\documentclass[%
 reprint,
superscriptaddress,
 amsmath,amssymb,
 aps,
 prb,
]{revtex4-2}

\usepackage{graphicx}
\usepackage{dcolumn}
\usepackage{bm}
\usepackage{hyperref}

\usepackage{xcolor}

\begin{document}

\title{Nanobridge SQUIDs as multilevel memory elements}

\author{Davi A. D. Chaves}
    \email{davi@df.ufscar.br}
\affiliation{Departamento de Física, Universidade Federal de São Carlos, São Carlos 13565-905, SP, Brazil}
\affiliation{Quantum Solid-State Physics, Department of Physics and Astronomy, KU Leuven, Celestijnenlaan 200D, B-3001 Leuven, Belgium}

\author{Lukas Nulens}
\affiliation{Quantum Solid-State Physics, Department of Physics and Astronomy, KU Leuven, Celestijnenlaan 200D, B-3001 Leuven, Belgium}

\author{Heleen Dausy}
\affiliation{Quantum Solid-State Physics, Department of Physics and Astronomy, KU Leuven, Celestijnenlaan 200D, B-3001 Leuven, Belgium}

\author{Bart Raes}
\affiliation{Quantum Solid-State Physics, Department of Physics and Astronomy, KU Leuven, Celestijnenlaan 200D, B-3001 Leuven, Belgium}

\author{Donghua Yue}
\affiliation{Quantum Solid-State Physics, Department of Physics and Astronomy, KU Leuven, Celestijnenlaan 200D, B-3001 Leuven, Belgium}

\author{Wilson A. Ortiz}
\affiliation{Departamento de Física, Universidade Federal de São Carlos, São Carlos 13565-905, SP, Brazil}

\author{Maycon Motta}
\affiliation{Departamento de Física, Universidade Federal de São Carlos, São Carlos 13565-905, SP, Brazil}

\author{Margriet J. Van Bael}
\affiliation{Quantum Solid-State Physics, Department of Physics and Astronomy, KU Leuven, Celestijnenlaan 200D, B-3001 Leuven, Belgium}

\author{Joris Van de Vondel}
    \email{joris.vandevondel@kuleuven.be}
\affiliation{Quantum Solid-State Physics, Department of Physics and Astronomy, KU Leuven, Celestijnenlaan 200D, B-3001 Leuven, Belgium}

\date{\today}

\begin{abstract}
    With the development of novel computing schemes working at cryogenic temperatures, superconducting memory elements have become essential. In this context, superconducting quantum interference devices (SQUIDs) are promising candidates, as they may trap different discrete amounts of magnetic flux. We demonstrate that a field-assisted writing scheme allows such a device to operate as a multilevel memory by the readout of eight distinct vorticity states at zero magnetic field. We present an alternative mechanism based on single phase slips which allows to switch the vorticity state while preserving superconductivity. This mechanism provides a possibly deterministic channel for flux control in SQUID-based memories, under the condition that the field-dependent energy of different vorticity states are nearby.
\end{abstract}

\maketitle

\section{\label{sec:Intro}Introduction}
Multilevel memory systems are technologically appealing, for instance, due to an increase in storage density~\cite{rozenberg2004nonvolatile,waser2009redox,zahoor2020resistive}. In past years, alternative non-semiconducting materials were proposed as non-volatile multilevel memories in photonic~\cite{rios2015integrated} and antiferromagnetic~\cite{olejnik2017antiferromagnetic} systems. In superconductors, a yTron current combiner can be used for non-destructive current readout and is capable of differentiating between discrete magnetic flux values trapped in superconducting loops~\cite{mccaughan2016using}, finding applications in proposed binary~\cite{zhao2018compact,mccaughan2018kinetic} and multilevel~\cite{toomey2019bridging} memory elements.

Superconducting quantum interference devices (SQUIDs) are widely used as powerful magnetometers and are a driving force in the development of quantum technology \cite{tinkham2004introduction,clarke1989squid,foley2009nanosquids,granata2016nano,kjaergaard2020superconducting}. In a conventional SQUID, superconductor-insulator-superconductor junctions present a sinusoidal current-phase relationship (C$\Phi$R) leading to a periodic single-valued dependence between the device critical current $I_c$ and the applied magnetic field $B$. However, if at least one of the junctions is an elongated constriction, such as a nanobridge or a nanowire, the C$\Phi$R of the junction becomes linear~\cite{hasselbach2002micro,podd2007micro}. As a result,  the $I_c(B)$ oscillations become linear as well and the four line segments of $I_c(B)$, two in the positive and two in the negative current side, define a so-called vorticity diamond~\cite{murphy2017asymmetric,dausy2021impact}. 

Additionally, the superconducting phase across such elongated constrictions can exceed 2$\pi$~\cite{dausy2021impact}. As such, different solutions exist at one magnetic field value (overlap of the vorticity diamonds), each corresponding with a different winding number, $n_v$, of the superconducting phase across the SQUID. Moreover, each vorticity state is characterized by a particular energy and critical current value. Therefore, a critical current measurement is a reliable tool to distinguish between different vorticity states. Finally, a single or series of topological fluctuations will trigger a transition to a different vorticity state or to the normal state by locally suppressing the order parameter, causing the phase across that given region to change by an integer multiple of 2$\pi$~\cite{lau2001quantum,mooij2006superconducting,aref2012quantitative,murphy2013universal,belkin2015formation}. These events are known as phase slips.

The SQUID vorticity may be set to a predetermined value by the establishment of proper writing protocols~\cite{lefevre1992thermal,palomaki2006initializing,ilin2021supercurrent}. This implies that information can be stored in the SQUID's different vorticity states. Indeed, flux-based qubits and memory elements relying on phase-slip physics have already been proposed~\cite{mooij2005phase,mooij2006superconducting,chen2020miniaturization,takeshita2021high,ilin2021supercurrent,ligato2021preliminary}. In this context, an applied magnetic field affects the SQUID energy and is a natural choice of control parameter in a flux-based memory. By providing a straightforward manner of changing the trapped flux inside a superconducting ring, it allows changing vorticities via phase slips. Recently, the role played by the field-dependent SQUID energy in the phase slip phenomena has gained attention. It was demonstrated that applied transport currents flowing in SQUIDs subjected to constant magnetic fields can also induce a transition between different vorticity states \cite{nulens2022metastable}. This fact has been explored in proposed zero-field memory elements \cite{takeshita2021high,ilin2021supercurrent}, in which the key ingredient is the creation of an asymmetry in the superconducting properties of the SQUID junctions. For a fully superconducting device, we previously demonstrated that the kinetic induction of the junctions may be tuned via nanofabrication, allowing control over the $I_c(B)$ characteristics of the device~\cite{dausy2021impact}.

In this paper, we report on the feasibility of SQUIDs containing nanometric-long nanobridges as multilevel memory elements. We demonstrate that by defining a field-assisted writing protocol, a large number of vorticity states can be accessed and then read at zero applied field, effectively creating an eight level memory system. However, writing to a specific state relies on a probabilistic process that requires the vorticity to be frozen in a determined state as the device transitions from the normal to the superconducting state. This process is heavily influenced by the total SQUID energy, an ingredient overlooked in previous studies. Using this information, we demonstrate that at applied fields for which adjacent vorticity states have equal energy, there is a very high probability of changing vorticity via single phase slips. By not triggering an unwanted transition out of the superconducting state, this mechanism constitutes an alternative and possibly deterministic channel for flux control.

\section{\label{sec:Sample}Experimental details}

\begin{figure}[htpb]
\includegraphics[width=\columnwidth]{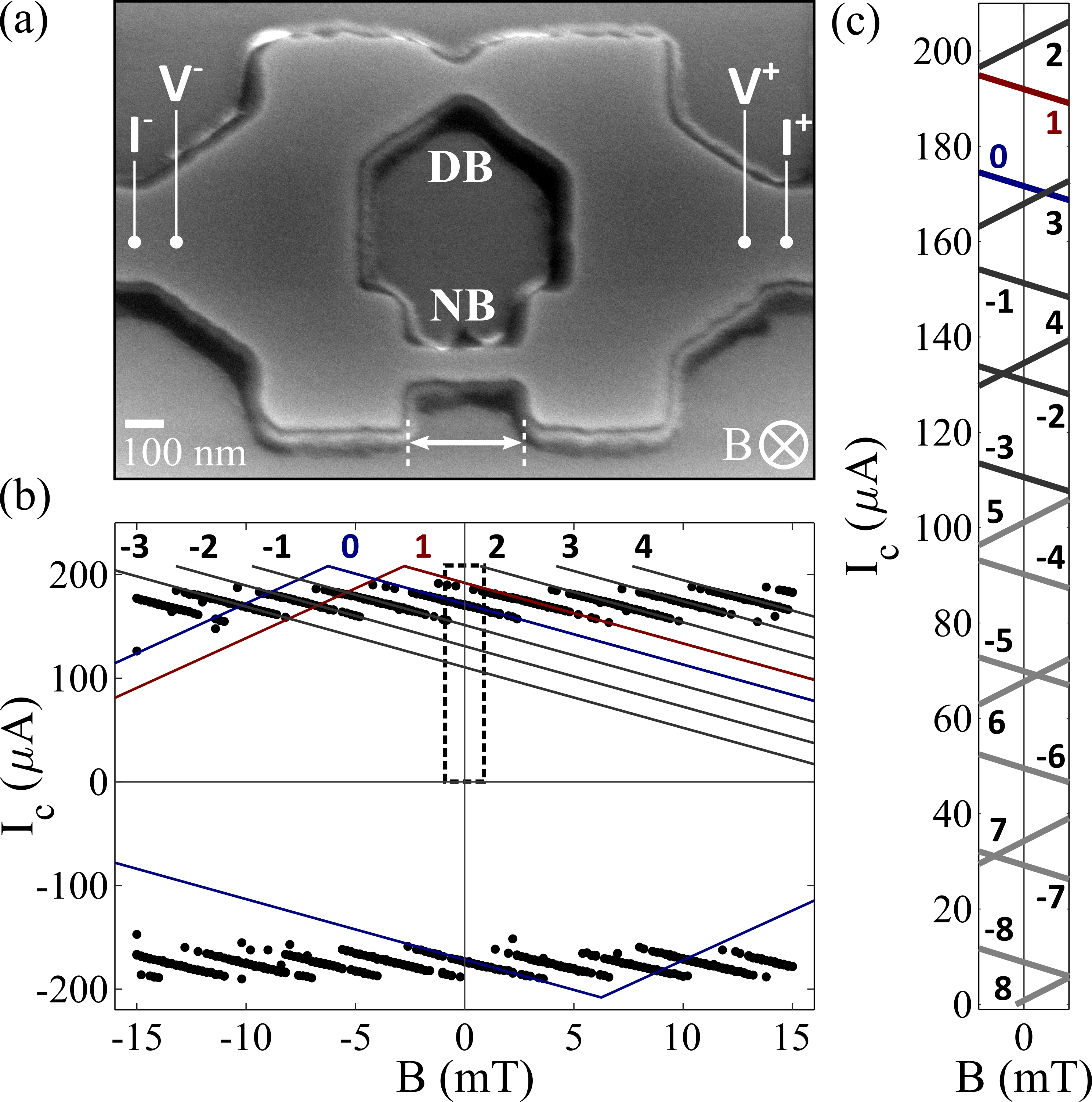}
\caption{\label{fig:1} (a) SEM image of a representative asymmetric Mo$_{78}$Ge$_{22}$/Au SQUID with nominal nanobridge length of 300~nm and width of 80~nm. (b) Characteristic $I_c(B)$ distribution for Device 400a. The points are gathered during three different sweeps for each magnetic field value and for each current direction. Numbered lines are fits of the vorticity diamond branches -- the Dayem bridge branch is shown only for $n_v$~=~0 and 1 for simplicity. (c) The vorticity diamonds for $-8~\leq~n_v~\leq~8$ in the region around 0~mT defined by the dashed rectangle in panel (b).}
\end{figure}

The samples investigated in this work are asymmetric Au-caped Mo$_{78}$Ge$_{22}$ nanosquids comprised of one Dayem bridge (DB) and one 40~nm-wide nanobridge (NB) of lengths varying from 100 to 400~nm. Figure~\ref{fig:1}(a) shows the SEM image of a prototypical device. To illustrate our discussion, we report on three selected samples, labeled Devices 250, 400a, and 400b according to the nominal length of the nanobrige. Devices 400a and 400b share the same geometry. By changing the NB length, we effectively controlled the superconducting properties of the junctions~\cite{dausy2021impact}. The devices were patterned using standard e-beam lithography. A 22~nm-thick Mo$_{78}$Ge$_{22}$ layer was deposited in a high-vacuum pulsed laser deposition chamber and the resulting structures were capped with a 3~nm thick gold layer to prevent oxidation. Critical current values were extracted from standard {\it VI} measurements using the four-point probe method, as shown schematically in Figure~\ref{fig:1}(a). During experiments, samples were kept in an Oxford Instruments Heliox $^3$He cryostat at a base temperature of 0.3~K. A room temperature $\pi$ filter with a cutoff frequency of 1~MHz and a Stanford Research Systems SR560 low-noise preamplifier were used in the measurements.

\section{\label{sec:Multilevel}Field-assisted writing: the SQUID as a multilevel memory element}

Initial $I_c(B)$ characterizations for Device 400a were performed as presented in Figure~\ref{fig:1}(b). These measurements, collected in three independent runs, allow the identification of different vorticity states based on their corresponding critical current values. The fit of $I_c(B)$ was performed following Equation~\eqref{eq:VortDiamond} and is represented by the numbered lines for different $n_v$ states in Figure~\ref{fig:1} (b)~\footnote{A direct inspection of the measured $I_c(B)$ distribution for Device 400a does not reveal the full shape of the vorticity diamond used to describe the critical current oscillation as only one of the branches appears. For this reason, we employed the preparation protocol explained in Section~\ref{sec:Multilevel} to obtain data relative to both diamond branches for Device 400a.}. As defined in Appendix~\ref{app:energy}, the results for Device 400a reveal values of $\Delta B~=$~3.5~$\pm$~0.1~mT, $I_c^{NB}$~=~101.5~$\pm$~0.3~$\mu$A, $I_c^{DB}$~=~106.7~$\pm$~0.6~$\mu$A, $\varphi_c^{NB}$~=~31.3~$\pm$~0.4~rad, $\varphi_c^{DB}$~=~20.1~$\pm.$~0.6~rad, $L_k^{NB}$~=~102~$\pm$~1~pH, and $L_k^{DB}$~=~62~$\pm$~2~pH, indicating a significant difference in the properties of the junctions.

These parameters allow us to trace the complete vorticity diamonds for Device 400a, extrapolating the regions in the $I_c(B)$ distribution where critical current measurements of a specific state are favored. For simplicity, Figure~\ref{fig:1}(b) only shows the four diamond branches for $n_v~=~0$. The choice for a very long nanobridge is based on the knowledge that such a constriction would present large critical phase difference and kinetic inductance~\cite{dausy2021impact}. As a consequence, a large number of states is expected to be found at a given applied field. Figure~\ref{fig:1}(c) highlights the region delimited by the dashed rectangle in panel (b) showing the vorticity diamonds for $-8~\leq~n_v~\leq~8$ around 0~mT. We observe that, although our initial experimental data reveals only two states at zero field, 17 total vorticities are expected to exist on the positive current side of the $I_c(B)$ distribution, supporting our design choices. To explore this rich landscape two things are paramount: the ability to write, i.e. select, the device vorticity state $n_v$; and the possibility to perform a readout of that state in a determined field value.

\begin{figure}[htpb]
\includegraphics[width=\columnwidth]{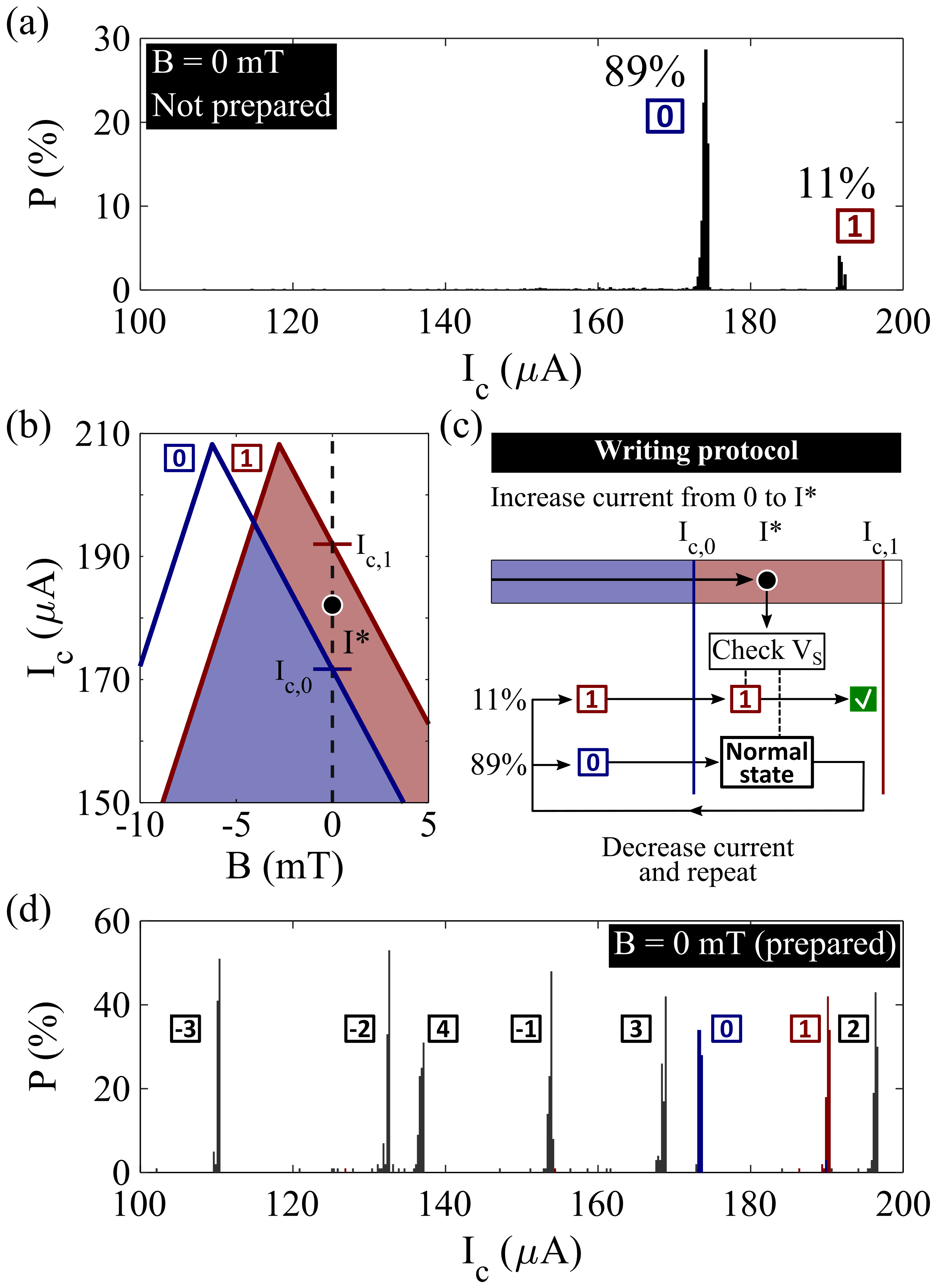}
\caption{\label{fig:2} (a) Critical current distribution for Device 400a after 1500 measurements at zero magnetic field when the vorticity state is frozen-in during a cool down from $+$220~$\mu$A, i.e., without preparing the device in any specific state. (b) Detail of the $I_c(B)$ distribution showing the vorticity diamond fit for $n_v~=~0$, blue solid line, and $n_v~=~1$, red solid line. The value of $I^*$ is chosen such that $I_{c,0}~<~I^*~<~I_{c,1}$. (c) A schematic representation of the writing protocol for vorticity state $n_v~=~1$. Repeatedly applying a current between 0 and $I^*$ will prepare the system in $n_v~=~1$. To gauge if the preparation is successful, the voltage $V_S$ across the device is checked with the current set at $I^*$. If $V_S$ reveals the SQUID is in the normal state, current should be reduced to zero, restoring the superconducting state, and the procedure repeated. If $V_S~=~0$, then the SQUID is in state $n_v~=~1$ and the preparation was successful. (d) $I_c$ distributions at zero magnetic field after preparing the device in different vorticity state.}
\end{figure}

The acquired $I_c(B)$ information can be used to implement a writing (or preparation) protocol that allows one to actively choose the device vorticity state \cite{murphy2017asymmetric,nulens2022metastable}. To illustrate this, consider Figure~\ref{fig:2}(a). It shows the critical current obtained for Device 400a after performing 1500 $I_c$ measurements at zero magnetic field. Comparing the results with the $I_c(B)$ distribution of Figure~\ref{fig:1}(b) we can only identify $I_c(0)$ values representing vorticity states $n_v$~=~1 ($I_{c,1}$~$\sim$~192~$\mu$A, with a probability of 11\%) and $n_v$~=~0 ($I_{c,0}$~$\sim$~172~$\mu$A, 89\%). These probabilities are dictated by the stochastic freeze-in process that determines the SQUID vorticity after entering the superconducting state at zero field. Notice that the highest $I_c$ state, $n_v~=~2$, was never observed. 
This is associated to the fact that, for sufficiently asymmetric SQUIDs~\cite{dausy2021impact,ilin2021supercurrent}, the highest $I_c$ state is not energetically favored when entering the superconducting state at zero magnetic field, as it presents a higher total energy than $n_v~=~0$ -- see Equation~\ref{eq:Energy}. Figure~\ref{fig:2}(b) then shows the $I_c(B)$ distribution around the positive peaks of $n_v~$=~0 and 1, highlighting their zero-field critical currents, $I_{c,0}$ and $I_{c,1}$. Figure~\ref{fig:2}(c) shows how a bias current $I^*$ may be used to gauge the SQUID vorticity. If $I^*~<~I_{c,0}$, as represented by the blue region, it is not possible to distinguish between $n_v~=~0$ and $n_v~=~1$, as both are possible at such bias current and the device will remain in the superconducting state regardless. However, if $I_{c,0}<I^*<I_{c,1}$ the device will remain superconducting if it is found in $n_v~=~1$, but will transition to the normal state if it was previously in $n_v~=~0$.
For the latter case, if the current is then reduced to zero, reestablishing superconductivity, the resulting vorticity state is dictated by the stochastic process with probabilities revealed by Figure~\ref{fig:2}(a). Therefore, the protocol to write Device 400a in $n_v~=~1$ consists of looping the current between zero and $I_{c,0}<I^*<I_{c,1}$ until the voltage $V_S$ across the device reads zero at $I^*$, i.e., it is in the superconducting state. The same principle is used in combination with a bias magnetic field, $B^*$, to successfully prepare the device in different states~\cite{nulens2022metastable}. 

\begin{table}[htpb]
\caption{\label{tab:PrepCoord}The bias current ($I^*$) and magnetic field ($B^*$) used in the field-assisted preparation protocol for Device 400a.}
\begin{ruledtabular}
\begin{tabular}{cccccccccc}
$n_v$          & -3  & -2  & -1   & 0   & 1   & 2   & 3   & 4   & 5   \\
$B^*$ (mT)     & -13 & -10 & -6.5   & -3.7 & 0   & 3.7 & 6.5 & 10 & 13.5  \\
$I^*$ ($\mu$A) & 178 & 178 & 178 & 180 & 180 & 183 & 183 & 183 & 183
\end{tabular}
\end{ruledtabular}
\end{table}

The use of our devices as memory elements requires the ability to reliably read different prepared states at the same applied magnetic field. To illustrate this, we prepared Device~400a in different vorticity states using the coordinates presented in Table~\ref{tab:PrepCoord}. By first reducing the bias current and then the bias field to zero, thus never leaving the constraints of the vorticity diamond, we are able to preserve the prepared states at zero magnetic field. Figure~\ref{fig:2}(d) reveals that the readout is successful at 0~mT for $-3~\leq~n_v~\leq~4$. It shows eight different histograms, each containing 100 $I_c$ measurements after preparing the device in the corresponding state before each current sweep. The observed critical currents are as low as $I_{c,-3}$~$\sim$~112~$\mu$A and as high as $I_{c,2}$~$\sim$~198~$\mu$A -- a variation of over 76\% and an enhancement of approximately 14.8\% for $I_{c,2}$ in comparison with $I_{c,0}$. The measurements in Figure~\ref{fig:2}(d) show close agreement with fits of the vorticity diamonds at zero field presented in Figure~\ref{fig:1}(c). State $n_v$~=~5, on its turn, could not be read at zero field, as the same 100 measurements yielded $I_c$ readings representing states $n_v~=~-3,-2$, and $-1$. This fact is related to Joule heating. For the previous eight states, with higher $I_c$, as the current was swept across the diamond edge, the heat dissipated by phase slips was large enough so that the device could not reestablish the superconducting state. For $n_v~=~5$, and all subsequent lower $I_c$ states, this is not the case. As the current reaches the diamond edge, a phase slip will occur without causing a cascade transition to the dissipative state. In other words, during the readout the vorticity is altered by a non-observable, or hidden phase slip. As such, no reliable reading is possible for these states at 0~mT. Figures~\ref{fig:2}(a)-(d) also reveal around 5\% of $I_c$ readings that cannot be traced to any specific $n_v$. These are random errors related to voltage fluctuations in the measurement setup that can be reduced by better line filtration.

This analysis reveals the ability of our SQUID to behave as an eight level memory, i.e., it presents eight different non-volatile states that can be read at a fixed magnetic field. Differently from a recently proposed memory element that relies on the physics of a single elongated weak-link \cite{ligato2021preliminary}, the writing between different states in Figure~\ref{fig:2}(d) is not a direct process, i.e., it does not require one sole physical stimulus like one magnetic field pulse. On the other hand, it is only by embedding the nanobridge in a SQUID that the possibility of exploring the physics of the weak-link to create a multilevel memory element arises. Moreover, the total number of available states may be tuned by modifications on the device design, as it is directly related to the slopes of the vorticity diamonds and to the vorticity state critical current, as shown in Figure~\ref{fig:1}(c). These are remarkable features that can only be accessed by the use of magnetic fields in the writing protocol and may be explored for sensing applications and in the development of non-volatile multilevel memory elements.

\section{\label{sec:DeterministicSlip}Towards non-probabilistic vorticity state switching}

The fact that the writing process depicted in Figure~\ref{fig:2}(c) is probabilistic is a major downside in comparison to alternative deterministic memory systems. Although a probabilistic writing scheme has been used previously in proposed memory elements~\cite{ilin2021supercurrent}, it is paramount to move towards a viable vorticity state switch protocol that does not rely on the process of leaving and returning to the superconducting state. Moreover, simply reducing the width of applied current pulses to manipulate the state does not solve the issue. As we demonstrated in Section~\ref{sec:Multilevel}, the frozen-in vorticity when the SQUID enters the superconducting state depends decisively on the device energy. In asymmetric SQUIDs~\cite{ilin2021supercurrent}, as our own, the lowest energy state may not present the highest $I_c$ at a desired applied field, hampering the writing process. 

\begin{figure}[h!]
\includegraphics[width=\columnwidth]{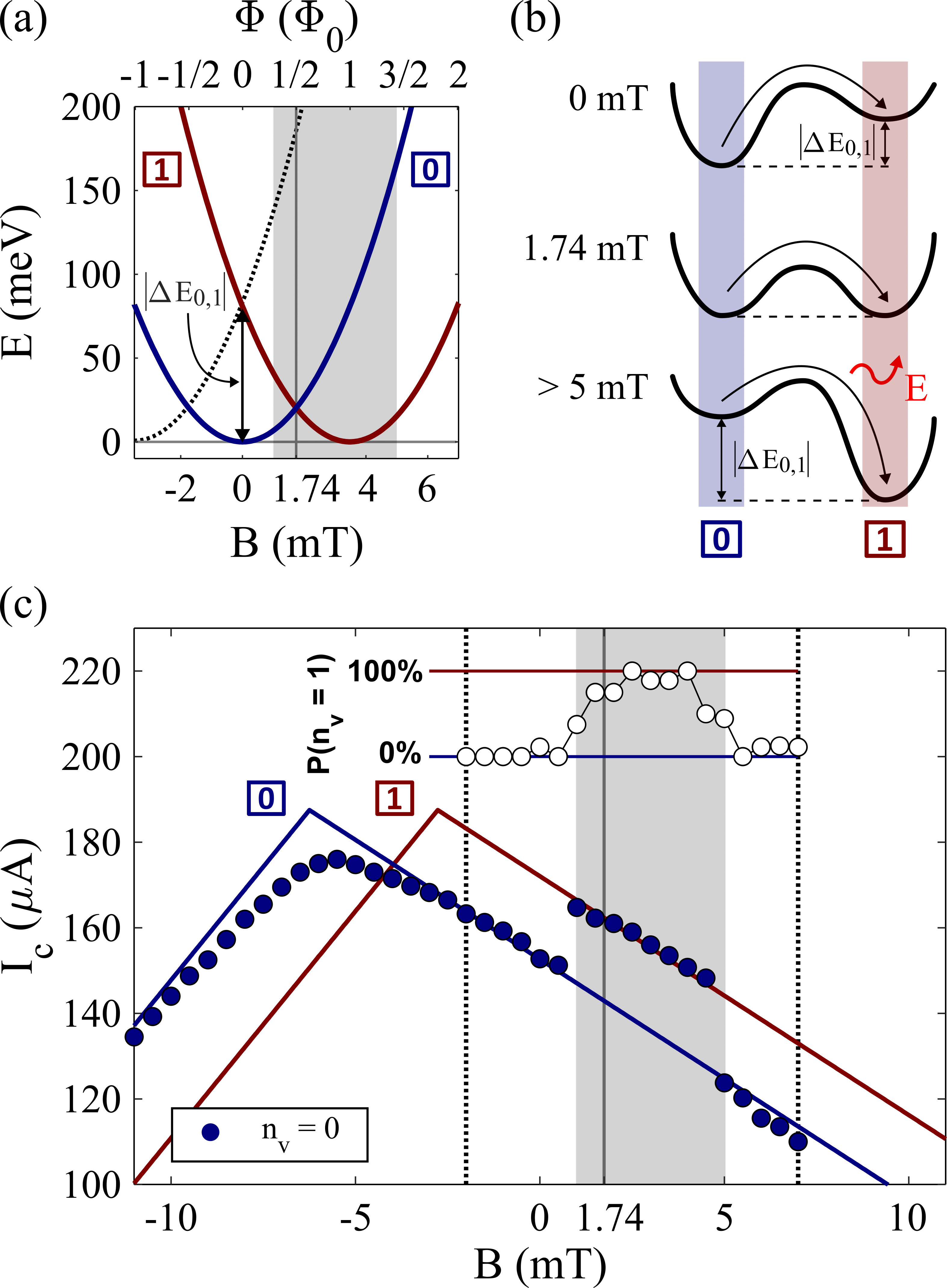}
\caption{\label{fig:3} (a) Energy versus magnetic field at zero current bias for $n_v$~=~0 and 1 for Device 400a obtained from the kinetic inductance. The dotted curve represents the energy for $n_v~=-1$. (b) Schematic representation of the energy landscape around the minima at $n_v$~=~0 and 1 at selected applied fields. The absolute energy difference $\Delta E_{0,1}$ is represented along a pictorial view of the energy barrier between the states. Curved dark arrows indicate the switch from state $n_v~=~0$ to 1 and the red arrow highlights the energy dissipated by one such phase slip at high applied fields. (c) Main panel shows one representative $I_c(B)$ measurement after preparing Device 400a in $n_v~=~0$. The respective vorticity diamonds for states $n_v$~=~0 and 1 are shown as solid lines. The inset shows the probability of finding the device in $n_v~=~1$ after a current sweep at particular field values as averaged for nine different measurements. In panels (a) and (c), a gray shaded area highlights the field values for which there is a high probability of observing single phase slips from $n_v$~=~0 to $n_v$~=~1.}
\end{figure}

Fortunately, it is possible to influence the probability of changing states without triggering irreversible heating to the dissipative state by exploring hidden phase slips -- if an applied magnetic field is used to favorably bias the energy~\cite{nulens2022metastable}. Figure~\ref{fig:3}(a) presents the energy versus magnetic field behavior at zero bias current obtained from the kinetic inductance values for Device 400a for $n_v$~=~0 and 1 according to Equation~\eqref{eq:Energy}. The field-dependent absolute energy difference between the states, labeled as $|\Delta E_{0,1}|$, is highlighted in the panel. At zero field, the energy values $E_0^{n_v}$ increase parabolically with $n_v$. As such, $E_0^{0}~=~0$ represents the more energetically favorable configuration for the investigated SQUID, which helps to explains why the probability that Device 400a freezes in $n_v~=~0$ is the highest when entering the superconducting state at zero field. Accordingly, the only other observed state, $n_v~=~1$, presents the second lowest energy at 0~mT and, therefore, the second highest probability. The possibility that a non-zero energy state appears in the measurements is related to random fluctuations experienced while the device enters the superconducting state~\cite{nulens2022metastable}. Figure~\ref{fig:3}(b) shows a schematic representation of the energy landscape around vorticity states $n_v$~=~0 and 1. Finally, although $E_0^{+1}~=~E_0^{-1}$, $n_v~=~-1$ is not observed in Figure~\ref{fig:2}(a). This happens because there is a dependence of the current polarity on the freeze-in probability, which is tied to the shape of the vorticity diamonds and can be attested in Figure~\ref{fig:1}(b) by noting that, for the negative current measurements, it is $n_v~=~-1$ and not $n_v~=~1$ that is observed at zero field.

Figure~\ref{fig:3}(c), meanwhile, reveals an intriguing feature. When the device is prepared in $n_v~=~0$, there is a defined field region, highlighted by the gray shaded area, showing a predominant probability of observing a single phase slip changing the vorticity of the SQUID to $n_v~=~1$. This is demonstrated in the main panel by one representative field-dependent $I_c$ measurement in which Device 400a was prepared in $n_v~=~0$ prior to a current sweep at each field value -- solid blue points \footnote{Here, the measured critical current values differ from those presented in Figure~\ref{fig:1}(b) because the device was removed from the experimental setup for approximately one month, which slightly depleted its superconducting properties.}. The blue and red solid lines are fits of Equation~\ref{eq:VortDiamond} for the vorticity diamonds for states $n_v$~=~0 and 1, respectively. Considering the blue points, for the left-most field values, the observed critical currents correspond to that of the prepared state, $n_v~=~0$. As the field is increased towards and beyond the line at $B$~=~1.74~mT, at which $E_B^{+1}~=~E_B^{0}$, a current sweep may induce a single phase slip and the observed $I_c$ now correspond to that of $n_v~=~1$. For measurements at even higher fields, Device 400a no longer experiences single phase slips and again the readout corresponds to the state $n_v~=~0$. This experiment was repeated nine times in the field region between the two dotted lines in Figure~\ref{fig:3}(c). The averaged probability of finding Device 400a in state $n_v~=~1$ after the current sweep is plotted in the inset. It shows that, at particular field values, the device will mostly endure a single phase slip to $n_v~=~1$ instead of transitioning to the normal state.

Such an observation may be better understood with the help of Figure~\ref{fig:3}(b).
At zero field, there is a non-zero energy difference between states $n_v~=~0$ and $n_v~=~1$ for Device 400a. On top of that, switching from $n_v~=~0$ to $n_v~=~1$ requires the SQUID to transpose a characteristic energy barrier. Then, as the current is swept across the edges of the vorticity diamond of $n_v~=~0$ at 0~mT, a phase slip to $n_v~=~1$ requires an amount of energy that is sufficient to drive the SQUID to the dissipative state. However, as the magnetic field is increased, $|\Delta E_{0,1}|$ decreases. At an applied field corresponding to $\Phi_0/2$, or 1.74~mT, $|\Delta E_{0,1}|~=~0$. For $B >$ 1.74~mT, $n_v~=~1$ presents the lowest energy and is thus the favorable state, as seen in Figure~\ref{fig:3}(a). Inspection of Figure~\ref{fig:3}(c) reveals that around such an applied field, Device 400a will mainly be found at $n_v~=~1$ after the current sweep. This is now the case because the energy released by a phase slip is lower than that required to switch from $n_v~=~0$ to $n_v~=~1$ at 0~mT, and the SQUID may sustain it while remaining in the superconducting state. As the field is increased above 1.74~mT, $|\Delta E_{0,1}|$ increases, ultimately causing a phase slip from $n_v~=~0$ to $n_v~=~1$ to release enough energy to again drive the device out of the superconducting state. For that reason, the observed vorticity state switch from $n_v~=~0$ to $n_v~=~1$ is not present at fields higher than 5~mT in Figure~\ref{fig:3}(c).

A second SQUID containing a 400~nm-long NB, labeled Device 400b, was studied. Completely analogous results were found and are shown in Appendix~\ref{app:400b}. Moreover, for Device 400b, we observed that preparation in all states between $-2 \leq n_v \leq 2$ resulted in a similar behavior, i.e., there were contained field regions that favored phase slips increasing the vorticity of the SQUID by $+1$. This suggests an alternative, step-by-step, vorticity state switching mechanism with very high probabilities, possibly even deterministic. Based on single phase slips, such mechanism would also present the benefit of not requiring leaving the superconducting state, thus reducing dissipation during memory operation.  

Importantly, this mechanism appears to be unidirectional in $n_v$, as we were not able to observe the reverse process in phase slips that decreased the vorticity. We can understand this considering the $I_c(B)$ distribution in Figure~\ref{fig:1}(b). If Device~400a was initially prepared in $n_v~=~0$, one could expect a high probability to observe phase slips to $n_v~=-1$ at $-1.74$~mT, corresponding to $-\Phi_0/2$ -- see dotted curve in Figure~\ref{fig:3}(a). However, around such an applied field, $n_v=~-1$ has a lower $I_c$ than $n_v~=~0$. It is then impossible for the vorticity to change to $n_v=~-1$ after the current is swept across the positive side of the vorticity diamond of $n_v~=~0$ and the device will transition to the dissipative state. Moreover, at zero-field, such phase slips may compromise the readout of the prepared state, therefore an optimal device design is very important. As such, the exploration of the presented effect as an alternative, possibly deterministic, and tunable flux control mechanism is an enticing possibility for the development of superconducting memory elements. In this effort, we find that a long NB is a necessity, as measurements conducted for devices containing 250~nm (see Appendix~\ref{app:400b}) and 300~nm NBs did not present particular regions with high phase-slip probabilities.

\section{\label{sec:Conclusions}Conclusions}

We fabricated different fully superconducting Mo$_{78}$Ge$_{22}$ SQUIDs comprised of one Dayem bridge and one nanobridge. Due to the tailored properties of the junctions, multiple vorticity diamonds intersect the zero-field line in the $I_c(B)$ distribution of our devices, which can be further influenced by the modification of the nanobridge length, an effective control parameter of the superconducting properties of the junctions. The implementation of a field-assisted writing protocol allowed reliable access to eight different vorticity states with distinct macroscopic properties at zero applied magnetic field, revealing the potential of our device as a superconducting multilevel memory element. An alternative approach to effectively switch the SQUID vorticity state by one winding number is presented, as for long enough nanobridges we demonstrate the existence of particular field regions where a current sweep favorably induces a single phase slip to the next vorticity state. Our analysis points to the determinant role played by the SQUID energy in the mechanisms involved in the freeze-in and phase slip probabilities, indicating applied magnetic fields as a vital ingredient in the effective manipulation of the vorticity states.

\section*{Acknowledgments}

This work was supported by Coordenação de Aperfeiçoamento de Pessoal de Nível Superior – Brasil (CAPES) - Finance Code 001, the São Paulo Research Foundation (FAPESP, Grant 2021/08781-8), the National Council for Scientific and Technological Development (CNPq, Grants 316602/2021-3, and 309928/2018-4), the China Scholarship Council (No. 202004890002), Research Foundation Flanders (FWO, Belgium, Grants G0A0619N, G0D5619N, and 11K6523N), and by COST action SUPERQUMAP
(Grant CA21144). D.A.D.C acknowledges Capes-PDSE grant number 88881.624531/2021-01.

\appendix

\section{Vorticity diamonds and energy}
\label{app:energy}

To obtain reliable readings of the SQUID vorticity states, it is necessary to have prior knowledge of their respective critical currents for different applied fields. For the SQUIDs employed in this work, we may consider the junctions to present linear C$\Phi$Rs of the form $I^{NB,DB}~=~I_{c}^{NB,DB}\left(\varphi^{NB,DB}/\varphi_c^{NB,DB}\right)$, where $I^{NB,DB}$ is the supercurrent and $\varphi_c^{NB,DB}$ the critical phase difference across the nanobridge and the Dayem bridge, respectively~\cite{dausy2021impact}. The left and right-side linear segments of the positive current side of $I_c(B)$ are then given by
\begin{equation}
    \label{eq:VortDiamond}
    \begin{aligned}
        I_c^{\text{left}}(B)~&=~I_{c}^{NB}~+ \\
        & +~\left(\frac{\Phi_0}{2\pi L_k^{DB}}\right)\left(\varphi_c^{NB}+2\pi\frac{B}{\Delta B}\right) \\
        I_c^{\text{right}}(B)~&=~I_{c}^{DB}~+ \\
        & +~\left(\frac{\Phi_0}{2\pi L_k^{NB}}\right)\left(\varphi_c^{DB}-2\pi\frac{B}{\Delta B}\right) \\
        \end{aligned}
\end{equation}
where $\Phi_0$ is the magnetic flux quantum, $L_k^{NB,DB}$ the kinetic inductance of the junction, and $\Delta B$ is the Little-Parks oscillation period defined by the area of the SQUID loop as $\Delta B~=~\Phi_0/A_{\text{loop}}$.

The ability to write the device to a specific state is a process related to the SQUID energy, as a freeze-in from the normal to the superconducting state will favor stable (less energetic) states over metastable ones~\cite{nulens2022metastable}. Then, by neglecting the self-inductance contribution estimated to be around 30 to 50 times smaller for a loop with the dimensions of our devices~\cite{dausy2021impact}, the total energy stored in the SQUID depends on the the bias current $I_{\text{bias}}$ and the magnetic field as \cite{nulens2022metastable}
\begin{equation}
    \label{eq:Energy}
    \begin{aligned}
        E_{B}^{n_v}~=&~\frac{1}{2}\left(\frac{L_k^{NB}L_k^{DB}}{L_k^{NB}+L_k^{DB}}\right)I_{\text{bias}}^2~+ \\ 
           & +~\frac{1}{2}\frac{1}{L_k^{NB}+L_k^{DB}}\left(\frac{B}{\Delta B}\Phi_0 -\Phi_0 n_v \right)^2
    \end{aligned}
\end{equation}
where $n_v$ is the vorticity state given by the relationship $\varphi^{NB} - \varphi^{DB} + 2\pi B/\Delta B = 2\pi n_v$, which represents the single-valuedness of the superconducting order parameter \cite{dausy2021impact}.

\section{Different devices}\label{app:400b}

\begin{figure}[h!]
\includegraphics[width=\columnwidth]{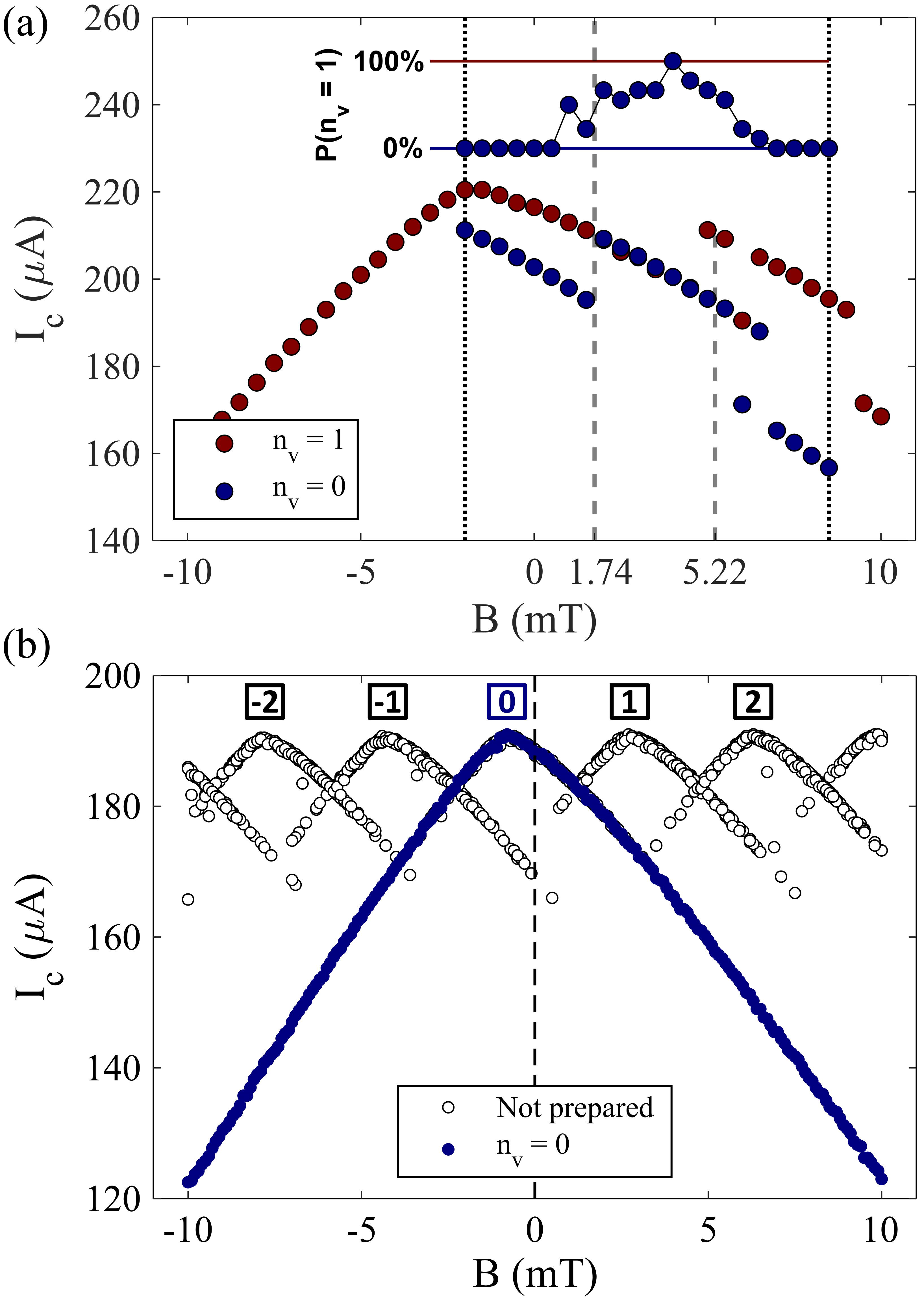}
\caption{\label{fig:4} (a) Main panel shows representative $I_c(B)$ measurements after preparing Device 400b in $n_v~=~0$ and  $n_v~=~1$. The inset shows the probability of finding Device 400b in $n_v~=~1$ after a current sweep at particular field values after it was initially prepared in $n_v$~=~0. The presented values are averaged from nine different measurements. (b) $I_c(B)$ for Device 250 when no preparation protocol is performed and after preparing in $n_v~=~0$ before each measurement.}
\end{figure}

Figure~\ref{fig:4} presents results obtained for devices different from the one discussed in the main text. Panel (a) demonstrates that, around $\Phi_0/2$, or 1.74~mT, Device~400b shows an increased probability of experiencing single phase slips from $n_v~=0$ to $n_v~=~1$. This confirms the findings described in Section~\ref{sec:DeterministicSlip} for Device~400a. Moreover, it also shows a completely analogous behavior at $3\Phi_0/2$, or 5.22~mT, at which field $|\Delta E_{1,2}|~=~0$. If Device~400b is prepared in $n_v~=~1$, it presents a high probability of experiencing a single phase slip to $n_v~=~2$. Panel (b) presents data for a device containing a shorter, 250~nm NB. No phase slips are observed as the current is swept across the vorticity diamond of the prepared state $n_v~=~0$. Therefore, we demonstrate that a long nanobridge is needed to create single phase slip-favorable field regions, as discussed in Section~\ref{sec:DeterministicSlip}.

\newpage

\bibliography{references.bib}

\end{document}